\documentclass[global,twocolumn]{svjour}

\usepackage{graphics}
\usepackage{graphicx}			
\usepackage{dcolumn}			
\usepackage{bm}					
\usepackage{xcolor}
\usepackage[utf8]{inputenc}
\usepackage[T1]{fontenc}
\usepackage{mathptmx}
\usepackage{etoolbox}
\usepackage{upgreek}            

\journalname{Appl. Phys. B}

\begin{document}

\title{Progress towards a matter wave interferometer for inertial sensing\\
with non-destructive monitoring of Bloch oscillations}

\author{D. Rivero \inst{1}
    \and C. Beli Silva \inst{1,2}
    \and M.A. Moreno Armijo \inst{1}
    \and H.~Ke\ss ler \inst{1,3}
    \and H.F.~da Silva \inst{1}
    \and G.~Comito \inst{4}
    \and R.F.~Shiozaki \inst{4}
    \and R.C.~Teixeira \inst{1,4}
    \and Ph.W.~Courteille \inst{1}}

\institute{Instituto de F\'isica de S\~ao Carlos, Universidade de S\~ao Paulo,
    13560-970 S\~ao Carlos, SP, Brazil
    \and Universiteit van Amsterdam, 1012 WX Amsterdam, The Netherlands
    \and Institut f\"ur Laser-Physik, Universit\"at Hamburg, 22761 Hamburg, Germany
    \and Departamento de F\'isica, Universidade Federal de S\~ao Carlos, 13565-905 S\~ao Carlos, SP, Brazil}
\date{Received: date / Revised version: date}

\maketitle

\begin{abstract}
We report on our progress in the construction of a continuous matter wave interferometer for inertial sensing via the non-destructive observation of Bloch oscillations. At the present stage of the experiment, around $10^5$ strontium-88 atoms are cooled down to below $1\upmu$K. Pumped by lasers red-tuned with respect to the $7.6~$kHz broad intercombination transition of strontium, the two counterpropagating modes of the ring cavity form a one-dimensional optical lattice in which the atoms, accelerated by gravity, will perform Bloch oscillations. The atomic motion can be monitored in real time via its impact on the counterpropagating light fields. We present the actual state of the experiment and characterize the laser spectrometer developed to drive the atom-cavity interaction.
\end{abstract}

\keywords{inertial sensing, Bloch oscillations, atom-cavity interaction, ultracold atom clouds}

\maketitle

\date{\today}

\section{Introduction}

Techniques for non-destructive real-time monitoring of the dynamics of a matter wave have great potential for improving atom interferometry. Applied, for example, to Bose-condensed atoms located in a standing light wave and subject to an external force, such techniques could record Bloch oscillations continuously with a single condensate. Consequently, the strength of the force can be measured in shorter times, the duty cycle can be reduced, and measurement uncertainties due to the process of generating new condensates can be avoided \cite{Peden09,Goldwin14,Samoylova14,Samoylova15,Patent15,Kessler16}.

Most imaging techniques monitor the trajectory of an atomic cloud by taking snapshots at different stages of its evolution via single shots of incident probe light. Unfortunately, the radiation pressure exerted by the probe light destroys the coherence of the matter wave, as the photonic recoil imparted by the scattered light is randomly distributed in all directions. This holds for techniques measuring the instantaneous density distribution (e.g.~absorption imaging), as well as for techniques measuring the velocity distribution, such as recoil-induced resonance spectroscopy \cite{Guo92} or Bragg spectroscopy \cite{Stenger99}. In fact, very few non-destructive techniques have been demonstrated so far. Dispersive imaging \cite{Andrews96} allows taking some dozens of pictures of a Bose-condensate before destruction. Electron beam imaging \cite{Ott08} is another example of non-invasive mapping of ultracold atomic density distributions.

A different approach makes use of optical cavities to steer the scattered light into a single cavity mode, making use of the very large Purcell factor of resonant cavities. Now, the scattering process becomes coherent, the mechanical impact of the incident light becomes predictable and can be taken into account, while heating can be avoided. The dynamics has been experimentally demonstrated in Refs.~\cite{Slama07,Bux11} using a ring cavity. In those experiments, when one of two counter-propagating cavity modes was pumped by a sufficiently far-detuned laser, a Bose-condensate confined in the mode volume responded by scattering light exclusively into the backward direction. The time-evolution of the recorded backscattered light contained all information on the condensate's trajectory, while the purely dispersive interaction with the cavity mode prevented decoherence of the condensate.

Eventually, atoms get lost due to parametric heating or collisions with the background gas. However, continuous matter wave interferometry has made tremendous progress with the recent achievement of continuous Bose-Einstein condensation \cite{Chen21}, and it seems very promising to combine methods allowing for a continuous refilling of the reservoir of atoms participating in the dynamics with non-destructive monitoring techniques.

The popularity of ultracold strontium \cite{Katori99,Yasuda04,Ido03,Takamoto05,Derevianko11} in matter wave interferometry \cite{Tino19,Rudolph20,HuLiang17,HuLiang20} has various reasons. The existence of a strong dipole-allowed transition (linewidth $\Gamma_{461}/2\pi=30.5~$MHz) and a narrow intercombination line (linewidth $\Gamma_{689}/2\pi=7.6~$kHz) allows rapid optical cooling close to the recoil temperature. The electronic ground state $^1S_0$ of the bosonic isotopes has no magnetic moment, which makes it insensitive to stray magnetic fields. The abundant $^{88}$Sr isotope exhibits a small $s$-wave scattering length ($a_s=-2a_B$, with $a_B$ the Bohr radius), such that interatomic collisions can be neglected. It is thus not surprising that very stable Bloch oscillations induced by gravity on ultracold $^{88}$Sr trapped in a vertical standing wave could be observed \cite{Ferrari06} and applied to gravity measurement. Finally, as we will show in this paper, the narrowness of the intercombination line, although requiring stable laser sources to drive it, facilitates cavity-assisted spectroscopy on this transition.

Combining the aforementioned advantages of $^{88}$Sr atoms with the coherent interaction between the atoms and the counter-propagating modes of a ring cavity will open up the path to a non-destructive, continuous measurement of gravity with state-of-the-art precision, as already proposed in literature \cite{Samoylova14,Samoylova15}. For this, an ultracold $^{88}$Sr cloud needs to be put in strong interaction with a ring cavity, quasi-resonant with the intercombination $^{88}$Sr transition and with a narrow laser source. Several technological requirements need to be fulfilled for achieving this goal. In this paper, we will present a novel setup for controlling the coherent interaction between the internal and external degrees of freedom of cold $^{88}$Sr atoms and laser light stored in a mode of an optical ring cavity, that fulfills all requirements for implementing non-destructive measurement of gravity. This setup differs from most other strontium experiments in several aspects. (i) The strontium atoms are provided from a two-dimensional magneto-optical trap (2D-MOT) \cite{Nosske17} enhanced with bi-chromatic 2D-MOT beams. (ii) The science chamber contains an optical ring cavity, which is pumped with laser light tuned close to the narrow intercombination line, and with which the atoms will interact. The ring cavity allows for novel effects related to the coherent interaction between the external degrees of freedom of the atoms and the light, such as collective atomic recoil lasing (CARL) \cite{Slama07,Bonifacio94,Kruse03}, which is not possible in linear cavities. We demonstrate in this paper, how to achieve full control over the ring cavity resonance and the pump laser frequency with respect to the narrow intercombination line $^1S_0\rightarrow {^3P}_1$ of strontium atoms, allowing for an exploration of different regimes of coherent interaction between the atoms and light. In particular, we will focus on the aforementioned regime of interest for monitoring atomic Bloch oscillations via the light which is backscattered from the ring cavity \cite{Samoylova14,Samoylova15}. The design and the characterization of the setup will be presented in Sec.~II. In Sec.~III we explain our scheme for controlling the ring cavity resonance frequency and the pump laser frequency independently with respect to the strontium intercombination line. In Sec.~IV we summarize the main features of the setup.

\section{Experimental setup}

In order to keep the experiment simple, we decided to trap and cool the atomic cloud at the same location of space, where the cavity-atom interaction takes place. In this way, we avoid having to transfer the cold atomic cloud over long distances from the preparation to the science region. The challenge of this approach, however, is to design the experimental setup such as to ensure optical access to all 6 ports of the ring cavity, as well as to the cooling beams of the magneto-optical trap (MOT), which also come from 6 different directions of space and must have a minimum diameter to ensure sufficient velocity capture ranges. Furthermore, there must be a clear passage for the atomic beam feeding the MOT and for the laser beams used for imaging the atomic cloud.
\begin{figure}[t]
	\centerline{\includegraphics[width=8.7 truecm]{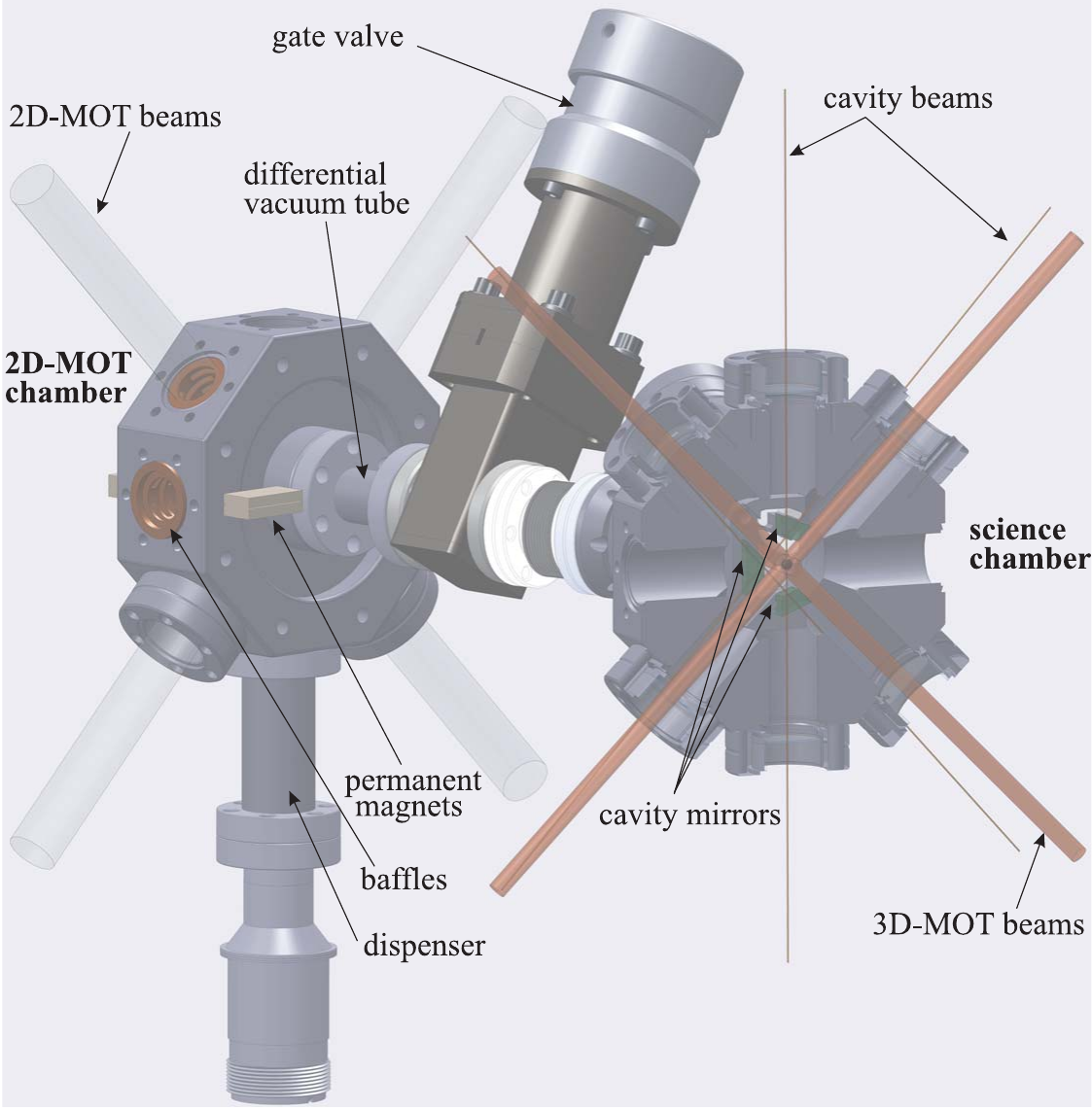}}
    \caption{(color online) Technical drawing of the 2D-MOT and the science chamber. 
        Note the ring cavity mounted inside the science chamber.}
    \label{fig:FigChamber}
\end{figure}

Fig.~\ref{fig:FigChamber} shows the layout of our experiment. The whole vacuum chamber setup fits on top of a 50 by $50~$cm base plate. It consists of two separate vacuum chambers linked by a $2~$cm long differential vacuum tube with $2~$mm inner diameter. Combinations of ion pumps and non-evaporable getter pumps allow us to maintain a vacuum of $10^{-8}~$mbar in the 2D-MOT chamber and $10^{-10}~$mbar in the science chamber.

The 2D-MOT is loaded from a strontium dispenser (AlfaVakuo e.U.) run at a current of $6.6~$A. We have observed that, although there is no direct path from the dispenser to the vacuum viewports and although the strontium atoms have the tendency to stick to the walls that they encounter, they eventually deposit on the viewports, coating them with an opaque layer. In order to minimize this problem, baffles have been mounted around the dispenser and the viewports of the 2D-MOT chamber obstructing simple atomic trajectories (such as single reflections from the walls). The baffles are simply stacks of concentric copper rings with an outer diameter made to fit into the CF16 ports of the vacuum chamber and an inner diameter leaving a $1~$cm clear aperture adapted to the size of the 2D-MOT beams. No coating of the viewports have been observed since this measure, which was implemented two years ago.

\begin{figure}[t]
	\centerline{\includegraphics[width=8.7 truecm]{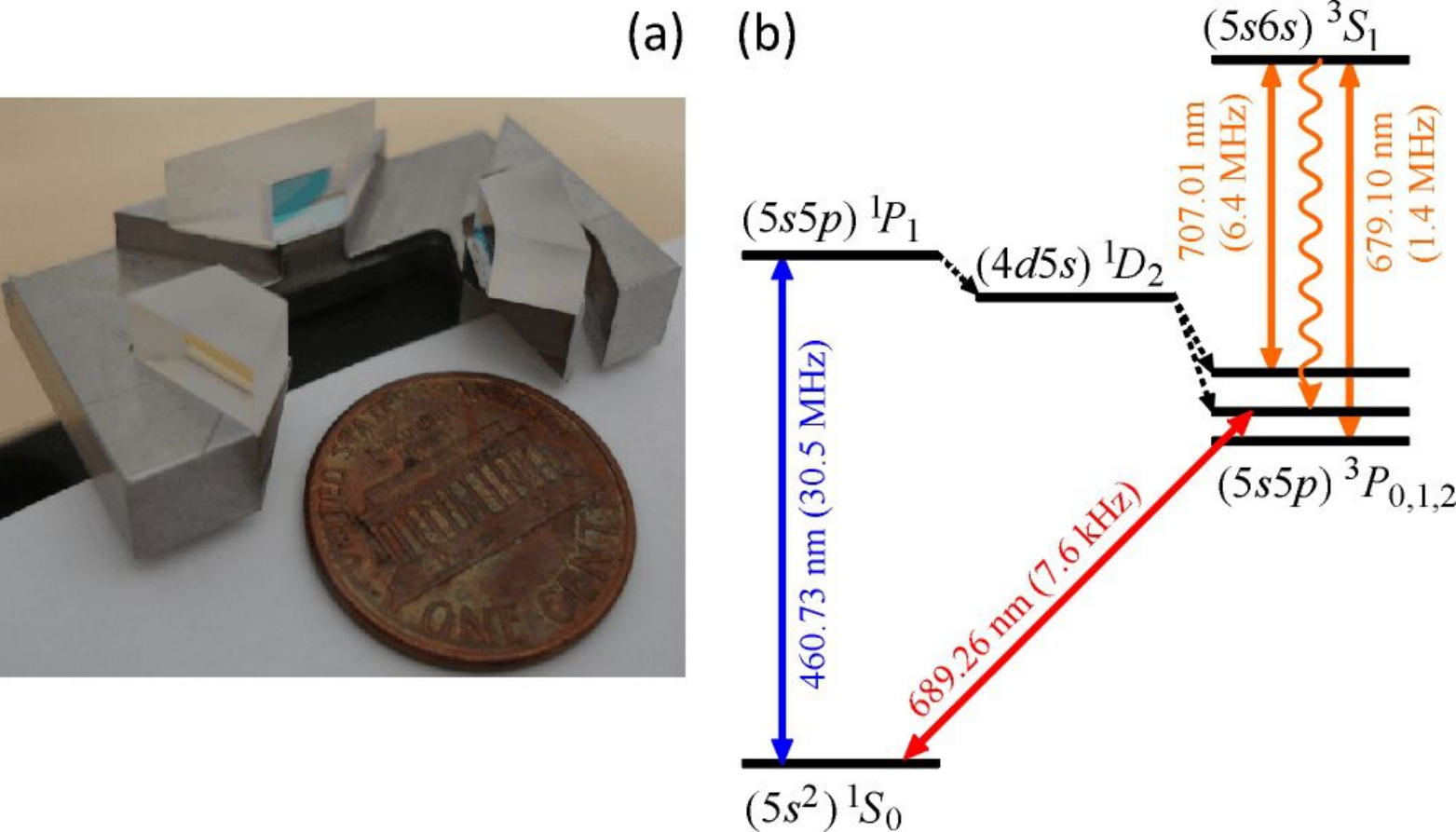}}
    \caption{(color online)(a) Picture of the ring cavity. The high reflecting mirror 
        at the right is mounted on a cantilever, which can be moved by a piezo transducer.
        (b) Level scheme of strontium showing the relevant transitions with their wavelengths and (in parentheses) the decay rates.}
    \label{fig:FigSrTransitions}
\end{figure}

The 2D-MOT is operated with permanent magnets arranged such that the symmetry axis is magnetic field-free \cite{Dieckmann98}. The retroreflected laser beams operating the 2D-MOT have $19~$mW each with waists of $0.7~$cm and are tuned $30~$MHz below the strong blue cooling transition $(5s^2) ^1S_0\leftrightarrow (5s5p)^1P_1$ at $461~$nm (see Fig.~\ref{fig:FigSrTransitions}). Additional single-path laser beams with $19~$mW power and tuned $134~$MHz below resonance are injected counterpropagating under $45^\circ$ to the atom beam ejected from the dispenser. These beams, which we will call 'slower beams' are meant to decelerate fast atoms and to increase the loading efficiency of the 2D-MOT \cite{Pedrozo-Penafiel16}.

The atoms captured in the 2D-MOT are illuminated by a resonant light beam, the so-called 'push beam', which has an intensity of $0.8I_{sat,461}$, where $I_{sat,461}=40.6~\textrm{mW/cm}^2$ is the saturation intensity of the cooling transition. The push beam accelerates the atoms towards the science chamber, where they are recaptured by a standard 3D magneto-optical trap called 'blue MOT'. The laser beams of the 2D-MOT and the blue MOT are generated from a frequency-doubled tapered-amplified diode laser (Toptica, DLC TA-SHG pro).

\subsection{Blue MOT} 

The blue MOT is realized with three pairs of counter-propagating laser beams, each beam with $3.5~$mW power tuned $32~$MHz below the blue cooling transition. Additionally, 'repumping' lasers are required to recycle the population of atoms eventually pumped into metastable states. One of them is the long-lived $(5s5p)^3P_2$ (see Fig.~\ref{fig:FigSrTransitions}(b)). We deplete this state by driving transitions at either $\lambda=707~$nm or at $\lambda=403~$nm \cite{Stellmer14,Moriya18}. However, both transitions can lead to non-negligible decay into another metastable state $(5s5p)^3P_0$, from which the atoms are recycled by driving a second repumping transition at $679~$nm (all lasers are Toptica, DLC pro). The joint action of the repumpers efficiently pumps the entire population from the metastable states to the $^3P_1$ state, from which the atoms finally decay into the ground state. The magnetic field for the blue MOT is created by a pair of coils in anti-Helmholtz configuration, being zero at the center of the coils and having a gradient of $65~$G/cm along the coils' axial direction.

\begin{figure}[t]
	\centerline{\includegraphics[width=8.7 cm]{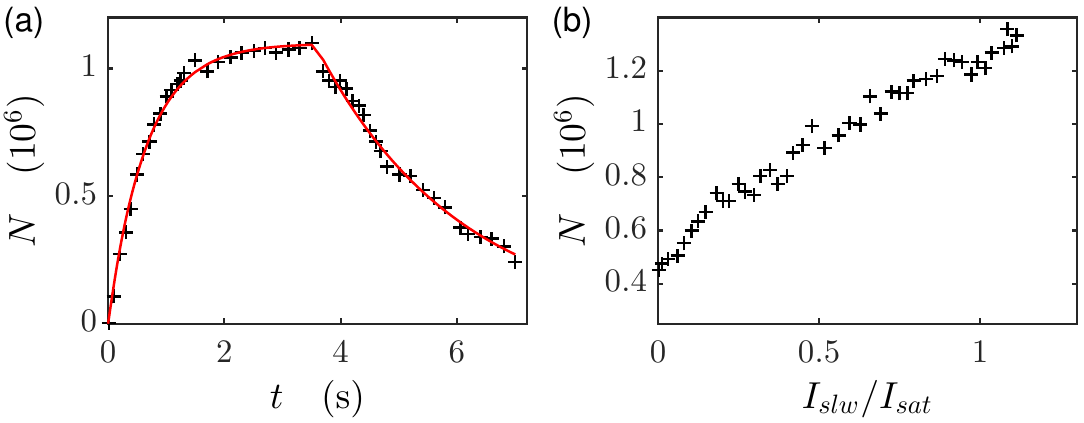}}
    \caption{(a) Blue MOT loading and decay after switching off the 2D-MOT push beam as a 
        function of time. The red curve is an exponential fit to the experimental data from which the time constants are extracted.
        (b) Number of atoms in the blue MOT as a function of the intensity of the slower beams of the 2D-MOT.}
    \label{fig:FigMotLoading}
\end{figure}

We currently trap around $N=10^6$ atoms at a temperature of $4~$mK in the blue MOT. While this atom number is relatively low in comparison to other experiments, and probably due to a conservative handling of the Sr-dispenser, it is plenty for the purpose of the intended experiments. Fig.~\ref{fig:FigMotLoading}(a) shows the number of atoms accumulating in the blue MOT as a function of time, with the light beams of the 2D-MOT turned on at $t=0$ and off at $t=3.5~$s, illustrating the loading and the decay of the blue MOT. We measure a loading time of $0.66~$s and a decay time of $2.47~$s. Fig.~\ref{fig:FigMotLoading}(b) shows the steady-state atom number of the blue MOT as a function of the intensity of the additional slower beams, as explained above.

\subsection{Red MOT} 

The temperature of the atoms cooled in the blue MOT is Doppler-limited by the linewidth of the transition to theoretically $T_D=\hbar\Gamma_{461}/2 k_B\approx 0.72~$mK, with $\hbar$ the reduced Planck constant and $k_B$ the Boltzmann constant. In practice, we reach $T_{blue}\approx 4~$mK, because of experimental imperfections. In order to reach lower temperatures, we cool the atoms in a second stage, called 'red MOT', operated on the narrow intercombination line, the $^1S_0\leftrightarrow{^3P}_1$ at $\lambda_{689} = 689~$nm, which is sufficiently narrow so that the Doppler-limit is below the recoil limit at $T_{rec}=\hbar^2k^2/2k_Bm\approx 460~$nK, with $k=2\pi/\lambda_{689}$ the wavenumber of the resonant light and $m$ the mass of a $^{88}\textrm{Sr}$ atom.

\begin{figure}[t]
	\centerline{\includegraphics[width=8.7 truecm]{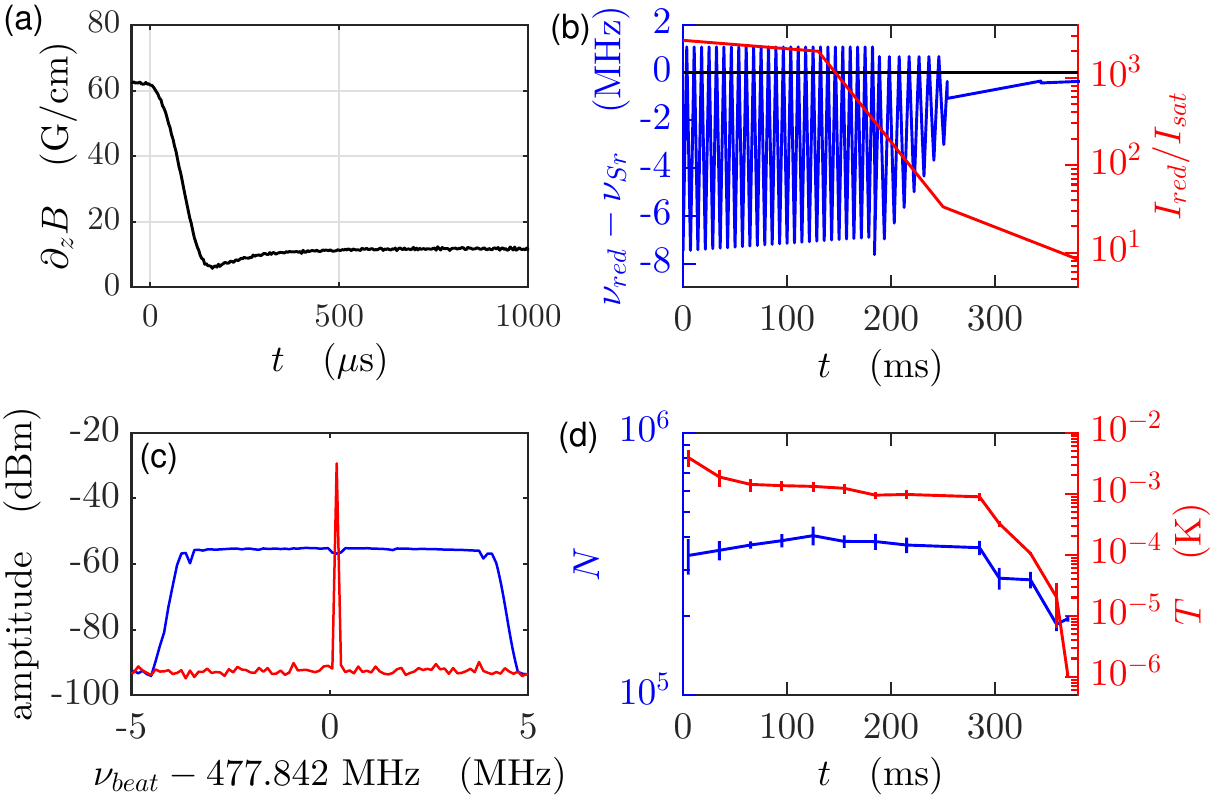}}
    \caption{(a) Switching of the magnetic field gradient (axial direction) between the blue and the red MOT.
		(b) Implemented red MOT frequency (blue) and power ramp (red). For clarity of illustration the modulation frequency, which in reality is $55~$kHz in the first part and $25~$kHz in the second, has been reduced by a factor of 200.
		(c) The blue curve shows the spectrum of the beat frequency between laser 1 ($\nu_{las1}$) and the red MOT light ($\nu_{redMOT}$) when the light injected into the slave laser is frequency-modulated with a triangular wavefunction with a fixed peak-to-peak frequency excursion of $16~$MHz (see main text). The red curve has been taken without this modulation.
		(d) Evolution of the measured atom number (blue curve) and temperature (red curve) along the modulation ramp.}
    \label{fig:FigMotSwitching}
\end{figure}

To transfer the atomic cloud from the blue to the red MOT, the magnetic field gradients are quickly switched within $160~\upmu$s from $65~$G/cm to $12.6~$G/cm, as shown in Fig.~\ref{fig:FigMotSwitching}(a). For this, we need to reduce the current in the anti-Helmholtz coils generating the magnetic fields from $8.75~$A to $1.7~$A. This is achieved with fast, high-voltage MOSFETs (type SCT2080KE, 1220V-44A) that allow for peak high voltage transients caused by the coils' inductance upon fast switching of their current, and a snubber circuit in parallel with the coil, composed of a capacitance of $47~\upmu$F in series with a resistance of $100~\Omega$.

The problem with using the intercombination line for a MOT is that its narrow linewidth entails a small velocity capture range, far below the Doppler width of the pre-cooled atomic cloud in the blue MOT. The temperature $T_{blue}$ of the atomic cloud at the time of the transfer from the blue to the red MOT corresponds to a half-width of the Doppler broadening of $\sqrt{k_BT_{blue}/m}/\lambda_{689}=1~$MHz at the wavelength of the intercombination line. This is much larger than the linewidth of the transition, which reduces the spectral overlap with the red MOT laser. In order to cool every velocity class of atoms within a certain velocity range, we frequency-modulate the red MOT laser with a $30~$kHz modulation frequency and a modulation excursion, which starts at $8~$MHz and is gradually reduced to 0 as the cloud cools down \cite{Katori99}. Simultaneously, the laser intensity of the red MOT is reduced from $2700I_{sat,689}$ to $8I_{sat,689}$ per beam, where $I_{sat,689}=3~\mu$W/cm$^2$ is the saturation intensity on this transition, in order to avoid excessive power broadening exceeding the Doppler width of the atomic cloud while it cools down. The frequency and the power ramps are illustrated in Fig.~\ref{fig:FigMotSwitching}(b). In Fig.~\ref{fig:FigMotSwitching}(c) we quantify the effective frequency-modulation of the red MOT light by beating it with the master laser 1 and recording the spectrum with a spectrum analyzer (Agilent, N9320B).

After $400~$ms of red MOT cooling, we typically have a cloud of $10^5$ atoms at a temperature of $800~$nK. As can be seen in Fig.~\ref{fig:FigMotSwitching}(d), we capture about 10\% of the atoms from the blue MOT. The final step consists in displacing the atomic cloud by applying an additional homogeneous magnetic field, until it overlaps with the mode of the ring cavity. Obviously, the stability of the red MOT laser frequency must be better than $\Gamma_{689}$. We will explain below our strategy to ensure this in practice. Observed shot-to-shot variations of the red MOT position are inferior to $10~\mu$m and thus do not represent a problem for the transfer into the $w=68.5~\mu$m waist of the ring cavity.

\subsection{Ring cavity characterization} 

A picture of our ring cavity is exhibited in Fig.~\ref{fig:FigSrTransitions}(a). It has the geometry of an isosceles right triangle, whose longer arm is aligned to the axis of gravity. It consists of a plane input coupler and two curved high-reflecting mirrors ($50~$mm radius of curvature) (Dorotek GmbH). The reflectivity of the input coupler is optimized for good impedance matching \cite{Comment01}, and we measure finesses of $F_s=1200$ for $s$-polarization and $F_p=500$ for $p$-polarization. The round trip length of the cavity is $3.64~$cm, which has been determined by measuring the free spectral range $\delta_{fsr}=c/L = 8.23~$GHz. From these data we calculate the parameters summarized in Tab.~\ref{table:1}, which characterize the ring cavity and the strength of the atom-cavity interaction.

The atomic cloud is positioned at the free-space waist of the ring cavity mode, where the mode diameter is $68.5~\mu$m. The cavity is in the 'bad cavity' limit, since its linewidth, $\kappa/\pi=\delta_{fsr}/F_s=6.86~$MHz, is much larger than both the recoil shift $\omega_{rec}=\hbar k^2/2m=(2\pi)4.78~$kHz and the transition linewidth, $\kappa\gg\omega_{rec},\Gamma_{689}$. The cavity-field coupling strength (single-photon Rabi frequency) can be calculated from $g_1=\sqrt{3\pi\Gamma\omega_0/k^3V}=(2\pi)8.7~$kHz, with $\omega_0=ck$ the angular frequency of resonant light and $V = 0.5~\textrm{mm}^3$ the Gaussian mode volume of the cavity, and from this we estimate the single-atom cooperativity of the cavity, $C=g_1^2/\kappa\Gamma$, and the the single-photon saturation parameter, $s=g_1^2/\Gamma^2$. Finally, the Table~\ref{table:1} lists the expected Bloch oscillation frequency $\nu_{blo}=mg/2\hbar k$, where $g$ is the gravitational acceleration.\\

\begin{table}
	\begin{tabular}[c]{lll}
		finesse	under $s$-polarization  & $F_s$				& $\approx 1200$\\
		finesse under $p$-polarization  & $F_p$				& $\approx 500$\\
		curvature of HR mirrors			& $\rho$			& $=50~$mm\\
		cavity round trip length		& $L$				& $=3.64~$cm\\
		cavity mode volume				& $V$				& $=0.5~mm^3$\\
		mode waist at atomic location 	& $w$				& $=68.5~\upmu m$\\
		free spectral range				& $\delta_{fsr}$	& $=8.23~$GHz\\
		atomic transition linewidth		& $\Gamma_{689}$    & $=(2\pi)7.6~$kHz\\
		cavity field decay rate         & $\kappa$		    & $=(2\pi)3.43~$MHz\\
		recoil shift					& $\omega_{rec}$	& $=(2\pi)4.78~$kHz\\
		coupling strength				& $g_1$				& $=(2\pi)8.7~$kHz\\
		number of atoms					& $N$				& $=10^5$\\
		lowest temperature achieved     & $T$				& $=800~$nK\\
		cooperativity					& $C$               & $=0.02$ \\ 
        1-photon saturation			    & $s$               & $=1.4$ \\ 
        Bloch frequency					& $\nu_{blo}$       & $=745~$Hz \\ 
	\end{tabular}\\
	\caption{Summary of parameters characterizing the ring cavity under $s$-polarization (except where stated otherwise), the atomic cloud trapped within the lowest-order Gaussian cavity mode, and the interaction between the atoms and light close to resonance with the $^1S_0\leftrightarrow{^3P}_1$ transition. Details in the main text.}
	\label{table:1}
\end{table}

\section{The red laser spectrometer}

\begin{figure*}[t]
	\centerline{\includegraphics[width=18 truecm]{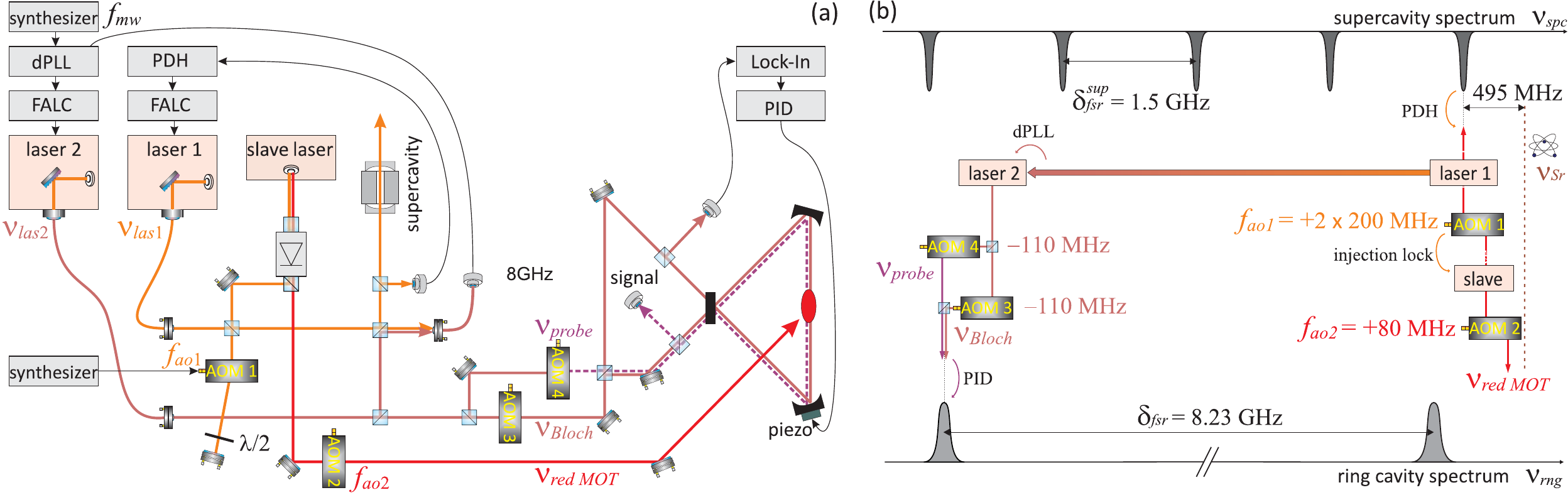}}
    \caption{(color online) (a) Optical setup of the red laser spectrometer. The acronyms
        are: dPLL: digital PLL, PDH: Pound-Drever-Hall stabilization unit, AOM: acusto-optic modulator, FALC: fast analog linewidth control (Toptica, FALC110), and PID: proportional-integral-differential servo (Toptica, DigiLock). The black arrow lines correspond to electronic signals, while the colored arrow lines correspond to laser beams, guided through polarization-maintaining fibers (curved lines) or propagating in free-space (straight lines). Different colors identify laser beams of different frequencies.
		(b) Locking scheme illustrating the frequency shifts and locking points of the lasers with respect to the supercavity spectrum (upper peak array) and the ring cavity spectrum (lower peak array). The $^{88}$Sr resonance $\nu_{Sr}$ is shown by a vertical dashed line.}
    \label{fig:FigRedSpectrometer}
\end{figure*}

The lasers operating near the $689~$nm intercombination line have three main purposes: (1) to cool the atoms down to temperatures close to the recoil temperature, (2) to generate the vertical standing wave along the ring cavity axis, in which the strontium atoms shall perform Bloch oscillations, and (3) to inject a probe light into the ring cavity for measuring the motional state of the matter wave. The optical setup of the red laser spectrometer is shown in Fig.~\ref{fig:FigRedSpectrometer}(a), and the locking scheme in Fig.~\ref{fig:FigRedSpectrometer}(b). For future experiments, we want to be able to control the length of the ring cavity with high precision, such that one of its resonance frequencies lies close to the strontium line or to the frequency of an independent laser pumping the ring cavity. 

We implement this via a sequential chain of three locking systems. A first laser (frequency $\nu_{las1}$) is locked to a stable reference cavity (supercavity), a second laser (frequency $\nu_{las2}$) is offset-locked to the first one via a digital phase-locked loop (dPLL), and finally the ring cavity length is stabilized to the second laser. We will explain the procedure in detail in the following sections.

\subsection{Providing near-resonant light for the red MOT} 

Near-resonant light is needed for the red MOT cooling stage and, eventually, for on-resonance excitation of the atoms. For that, we stabilize the laser 1, of frequency $\nu_{las1}$, to a mode of a supercavity (Stable Laser Systems, specified finesse 250000, measured finesse 185000) by a Pound-Drever-Hall locking scheme (PDH) \cite{Drever83}. The frequency gap between the strontium line ($\nu_{Sr}=434.829~121~311~$MHz \cite{Ferrari03}) and the nearest supercavity mode is measured to be $\nu_{Sr}-\nu_{sup}=477.84~$MHz. To overcome the gap, the laser beam is double-passed through a first AOM ($f_{ao1}$), amplified by injection into a slave laser which, after single-passing another AOM ($f_{ao2}$), finally illuminates the atomic cloud. The frequency modulation of the red MOT light (see above and Fig.~\ref{fig:FigMotLoading}(d)) is implemented by modulating the frequency $f_{ao1}/2$ of the RF produced by a synthesizer (Rohde \& Schwarz, SMB100A) controlling the double-pass AOM. The injection lock is clearly fast enough to handle the modulation, as demonstrated in Fig.~\ref{fig:FigMotSwitching}(c), which shows the presence of the frequency modulation in the light amplified by the slave laser.

\begin{figure}[b]
	\centerline{\includegraphics[width=8.7 truecm]{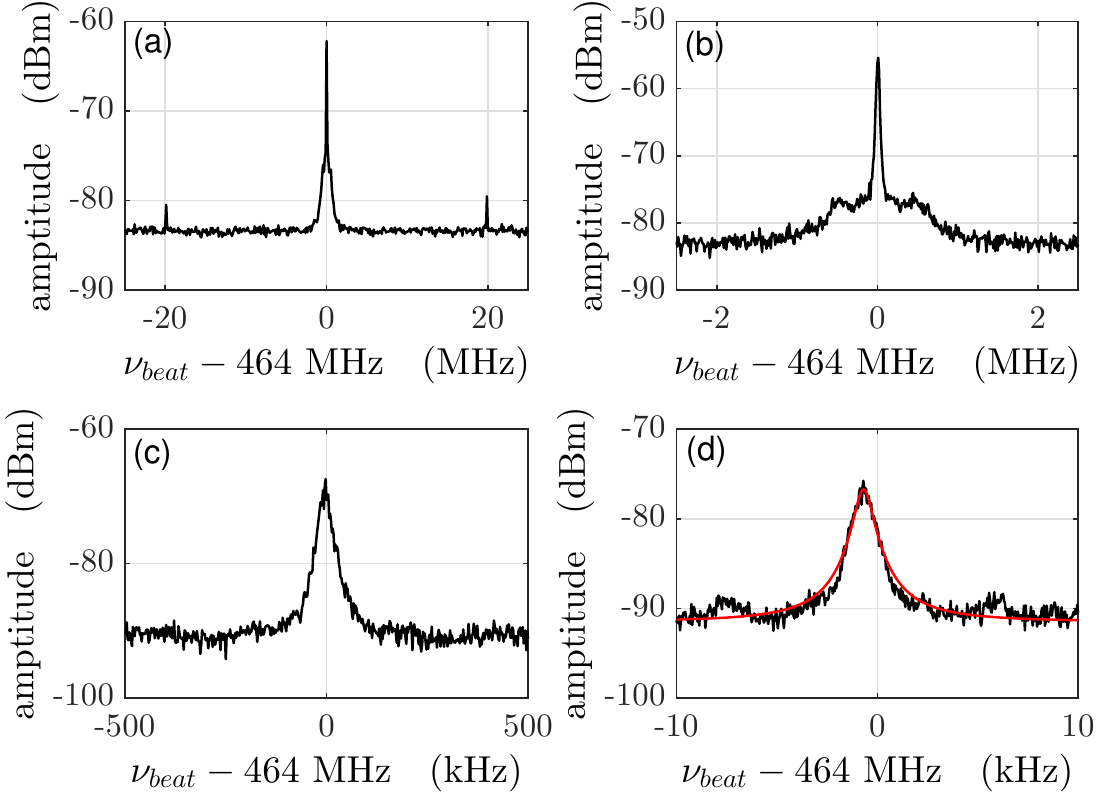}}
    \caption{Characterization of the performance of the Pound-Drever-Hall (PDH) stabilization. 
        (a-d) Beat signal of two independent PDH-locked lasers with $464~$MHz frequency difference recorded with various resolution bandwidths of the spectrum analyzer: (a,b)~$30~$kHz, (c)~$1~$kHz, and (d)~$10~$Hz. The PDH modulation sidebands are visible as small peaks at $\pm 20~$MHz. The red curve in (d) is a Lorentzian fit with a width of $450~$Hz. This allows us to assert that any one of the lasers has an emission bandwidth below this width.}
    \label{fig:FigSr1Sr2Beat}
\end{figure}

We characterized the frequency stability of our slave laser by beating it with an independent laser located in an adjacent laboratory and locked to a different reference cavity. The spectrum of the beat signal, centered around the frequency difference of $464~$MHz between both lasers, is shown in Fig.~\ref{fig:FigSr1Sr2Beat}(a-d) for different resolution bandwidths of the  spectrum analyser. As seen in Fig.~\ref{fig:FigSr1Sr2Beat}(d), the spectral width of the beat signal is $<1~$kHz, the measurement being limited by acoustic noise entering via the $10~$m long fiber guiding the laser light from one lab to the other. This demonstrates that the absolute emission bandwidth of both lasers is below $1~$kHz, which is sufficient for spectroscopy on the intercombination transition.

\subsection{Controlling the cavity length} 

\begin{figure}[t]
	\centerline{\includegraphics[width=8.7 truecm]{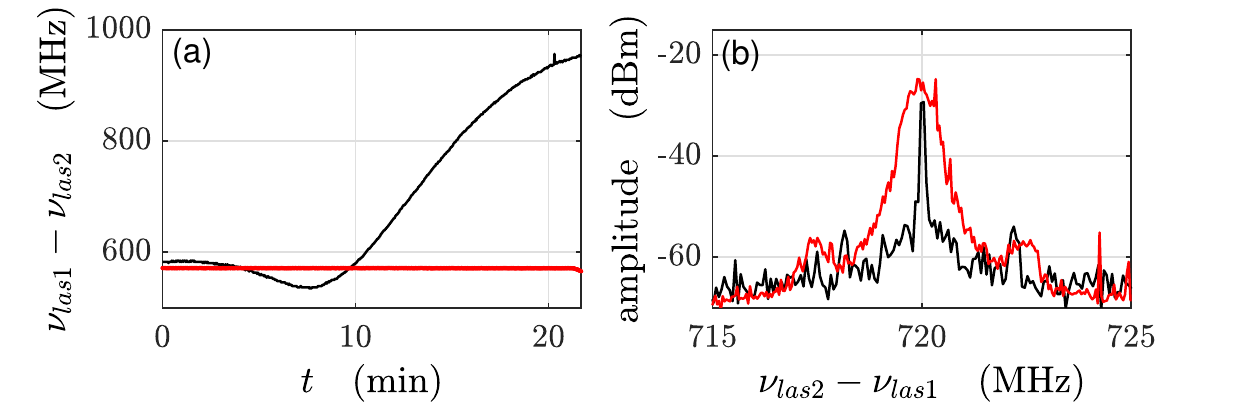}}
    \caption{(a) Long-term characterization of the frequency drifts of the ring cavity.
        Laser 2 was locked to the ring cavity and laser 1 to the supercavity. The curves were taken by beating both lasers and monitoring the beat frequency in time using a spectrum analyzer. The black curve shows the natural drifts of the ring cavity. The red curve shows the same signal when the ring cavity was additionally locked to laser 1 via a lock-in amplifier and a PID servo electronics (Toptica, DigiLock).
        (b) Short-term spectrum of the beating note for each measurement of (a). The black curve demonstrates the short-term stability of the ring cavity. The fluctuations come from acoustic noise afflicting the ring cavity. The small sidebands at $\pm 2~$MHz are due to the different PDH modulation frequencies for both lasers ($20~$MHz for laser 1 and $18~$MHz for laser 2). For the red curve, the ring cavity was additionally locked to laser 1. The spectral broadening in this measurement, which comes from the modulation of the ring cavity piezo, can be avoided by replacing the lock-in amplifier by a PDH servo.}
    \label{fig:FigRingcavityDrifts}
\end{figure}

The second laser ($\nu_{las2}$) can be locked to the ring cavity using the PDH technique in the same way as for the stabilization of laser 1. This allows us to monitor the temporal drift of the ring cavity length by measuring the frequency of laser 2 as a function of time. Fig.~\ref{fig:FigRingcavityDrifts}(a) shows the time evolution of the frequency beat between laser 2 (locked to the ring cavity) and laser 1 (locked to the stable supercavity), as monitored with a spectrum analyzer. The fluctuations of several $10~$MHz over minutes reflect the instability of the ring cavity, which is not mechanically nor thermally isolated. Fig.~\ref{fig:FigRingcavityDrifts}(b) shows the instantaneous beat spectrum of the two lasers. The width of the spectrum is a measure for the short term stability of the laser locked to the ring cavity and reflects the quality of the PID-lock.

The large frequency drifts observed for the ring cavity resonance frequency were expected and demonstrate the necessity of its stabilization. For this purpose, one of the cavity mirrors is mounted on a piezo-electric transducer (PZT) (Physik Instrumente, PICMA PL0XT0001). This allows us to tune the length of the ring cavity and to stabilize it to the frequency $\nu_{las2}$ of laser 2. In practice, we achieve this stabilization via a lock-in amplifier and a proportional-integral-differential (PID) servo electronics (Toptica, DigiLock).

Laser 2, in turn, can be locked to a stable reference. The locking scheme should, however, satisfy two conditions. First of all, it is important to avoid perturbing (heating) the atomic cloud stored inside the ring cavity by the laser field. This implies that the frequency $\nu_{las2}$ of laser 2 must be tuned sufficiently far enough from the strontium resonance. Thanks to the narrowness of the intercombination line, detuning the laser to the next adjacent ring cavity mode, which is $8.23~$GHz away, is sufficient to avoid spontaneous scattering processes. As an example, assuming $100~$mW of intracavity power, at this detuning the scattering rate is only on the order of $5~\textrm{s}^{-1}$ per atom. Second, we aim for wide tuning ranges covering the whole spectral range of the ring cavity. Both conditions are met by a digital PLL, as detailed in the next section.

\subsection{Phase-locking of the lasers $\nu_{las1}$ and $\nu_{las2}$} 

In our spectrometer, exhibited in Fig.~\ref{fig:FigRedSpectrometer}(a), we implement an offset-lock of laser 2 to the laser 1 by beating both lasers on a fast photodetector (Thorlabs, PDA8GS). The beat frequency is now divided by 32 via a digital PLL (Analog Devices evaluation board, EVAL-ADF4007) \cite{Appel09} and compared to the RF frequency $f_{mw}$ provided by a stable frequency synthesizer (Rohde \& Schwarz, SMB100A). The comparison generates an error signal, which is fed into the PID-electronics (Toptica, FALC110) controlling the laser (Toptica, DLpro). Varying $f_{mw}$, the laser frequency $\nu_{las2}$ can now be tuned with $1~$Hz precision, and through it the length of the ring cavity. As illustrated in Fig.~\ref{fig:FigRedSpectrometer}(b), the resonance frequencies of the ring cavity can be varied over wide ranges, e.g.~it is possible to tune a cavity mode adjacent of the mode locked to laser 2 across the strontium line.
\begin{figure}[t]
	\centerline{\includegraphics[width=8.7 truecm]{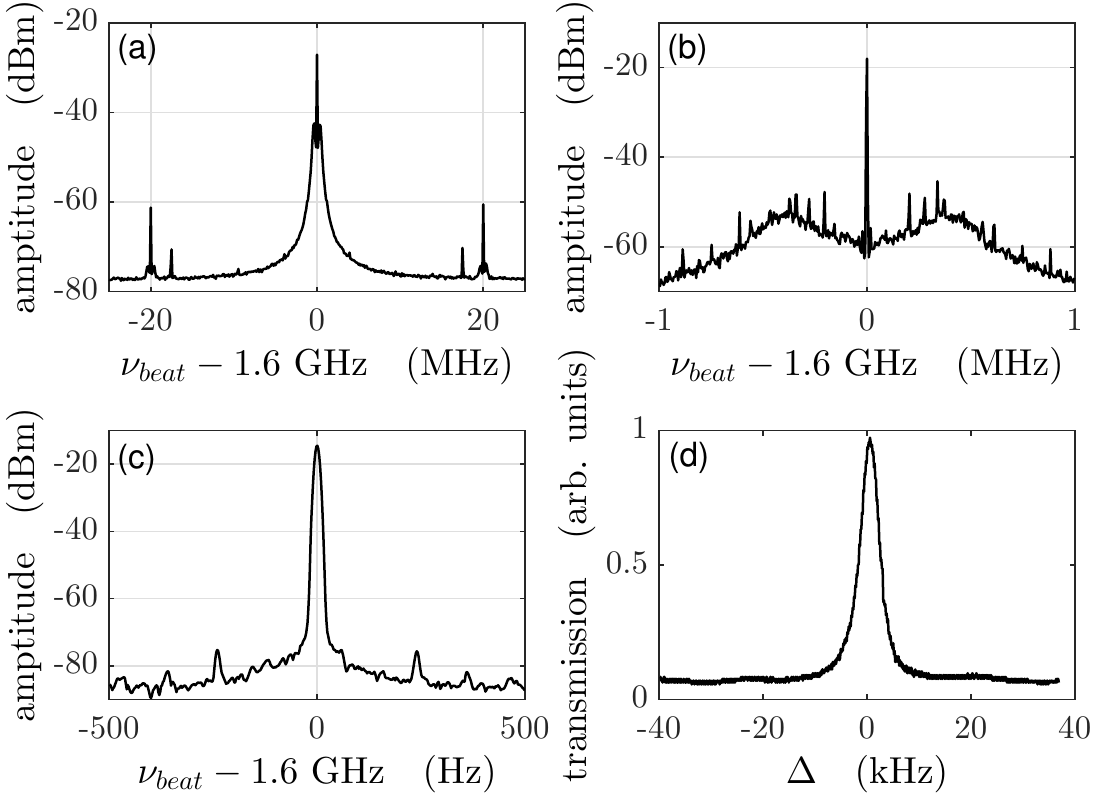}}
    \caption{Characterization of the performance of the digital PLL. One laser
        ($\nu_{las1}$) was locked to a mode of the supercavity by means of a Pound-Drever-Hall stabilization. The other laser ($\nu_{las2}$) was locked to the first one by means of a digital PLL, as described in Ref.~\cite{Appel09}.
		(a) Beat signal of the two mutually phase-locked lasers recorded with 30~kHz resolution bandwidth of the spectrum analyzer. For this measurement, the lasers had a frequency difference of 1.6~GHz.
		(b) Same as (a) but with 3~kHz resolution bandwidth. 
		(c) Same as (a) but with 10~Hz resolution bandwidth. 
		(d) Transmission spectrum of the supercavity. Laser $\nu_{las1}$ was locked to a cavity and generated a constant transmission visible as a small offset at all detunings. Additionally, laser $\nu_{las2}$ was ramped over an adjacent supercavity mode. The measurement yields the free spectral range of the supercavity $\delta_{fsr}^{sup}=1.445065~$GHz, and from the 5~kHz width of the spectrum we infer a finesse of $F=185000$.}
    \label{fig:FigDPLLBeat}
\end{figure}

We characterize the performance of the digital PLL stabilization in two ways: (i) Fig.~\ref{fig:FigDPLLBeat}(a-c) show beat frequency measurements between the (free-running) laser 1 and laser 2 being DPLL-locked to laser 1 for various resolution bandwidths. The $10~$Hz width of the peak in Fig.~\ref{fig:FigDPLLBeat}(c) is limited by the minimum resolution bandwidth of the spectrum analyzer (Agilent, N9320B) of $10~$Hz. (ii) We locked laser 1 to a supercavity mode via a PDH electronics, laser 2 on laser 1 via the digital PLL, and ramped the synthesizer frequency which feeds the PLL such as to tune laser 2 across an adjacent supercavity mode. The transmission spectrum of the supercavity is shown in Fig.~\ref{fig:FigDPLLBeat}(d). The center of the transmission peak allows us to extract the free spectral range of the supercavity with high precision, $\delta_{fsr}^{sup}=1.445065~$GHz. From that, and from the width of the transmission peak, we obtain a finesse of the supercavity of $F=185000$.

\section{Conclusion}

In this paper we described our approach to constructing a matter wave gravimeter with real-time monitoring of Bloch oscillations. The experimental setup is nearly completed: we are now able to create clouds of about $10^5$ strontium atoms with a temperature below $1~\upmu$K, which we can transfer into the mode volume of a ring cavity. The length of the ring cavity and the frequency of the narrow interrogation laser can also be tuned at will, providing all ingredients for implementing a non-destructive measurement of gravity. We will now start to search for signatures of atom-cavity interactions. First, we plan to demonstrate trapping of atoms by the light fields in the cavity tuned some $100~$MHz below resonance. Next, we will search for signatures of the presence of atoms in the light fields of the cavity modes, e.g.~due to a collective response of the atomic motion to incident light \cite{Kruse03} or by observing normal-mode splitting of cavity resonances \cite{Culver16}. Finally, we will search for signatures of Bloch oscillations in the light fields, as predicted in Refs.~\cite{Samoylova14,Samoylova15}, in order to measure gravity with our setup.

\section*{Acknowledgments}

The authors acknowledge the Brazilian agencies for financial support. H.K., and Ph.W.C. hold grants from S\~ao Paulo Research Foundation (FAPESP) (Grant Nos. 2016/16598-0, and 2013/04162-5, respectively). Furthermore, Ph.W.C. holds grants from CAPES (Grant Nos. 88881.1439362017-01 and 88887.13
01972017-01). D.R., C.B. and M.M. Thank the Coordenação de Aperfeiçoamento de Pessoal de Nível Superior – Brasil (CAPES) – Finance  Code 001 for the scholarships on which this study was partly financed.

\medskip
\noindent\textbf{Disclosures.} The authors declare no conflicts of interest. The data that support the findings of this study are available from the corresponding author upon reasonable request.


\begin{thebibliography}{99}
\bibitem{Comment01} Impedance matching is the optimization of the cavity resonant power enhancement by tuning the input coupler reflectivity for given transmission and absorption losses of all three cavity mirrors.
\end{thebibliography}
\end{document}